\documentclass[12pt]{article}
\usepackage{graphicx}
\usepackage{amsmath}
\input psfig.sty

\newlength{\back}
\newlength{\greater}

\begin{document}

\begin{titlepage}

\title{Spontaneous Symmetry Breaking\\ in the Space-time of an Arbitrary
Dimension}

\author{Je-An \ Gu\thanks{%
E-mail address: wyhwange@phys.ntu.edu.tw} \ \ \ and \ \ W-Y. P. Hwang\thanks{%
E-mail address: wyhwang@phys.ntu.edu.tw} \\
{\small Department of Physics, National Taiwan University, Taipei
106, Taiwan, R.O.C.}
\medskip
}

\date{\small \today}

\maketitle

\begin{abstract}
We propose a new scenario to implement spontaneous symmetry
breaking in the space-time of an arbitrary dimension ($D>2$) by
introducing the non-minimal coupling between the scalar field and
the gravity. In this scenario, the usage of the familiar $\lambda
\Phi ^{4}$ term, which is non-renormalizable for $D\geq 5$, can be
avoided altogether.
\end{abstract}

\vspace{6cm}

\begin{flushleft}
\footnotesize \emph{Keywords:} %
\footnotesize Spontaneous symmetry breaking; Non-minimal coupling;
Arbitrary dimension
\end{flushleft}

\end{titlepage}

\section{Introduction}

\begin{figure}[htb]
\begin{minipage}{60mm}
\centerline{\psfig{figure=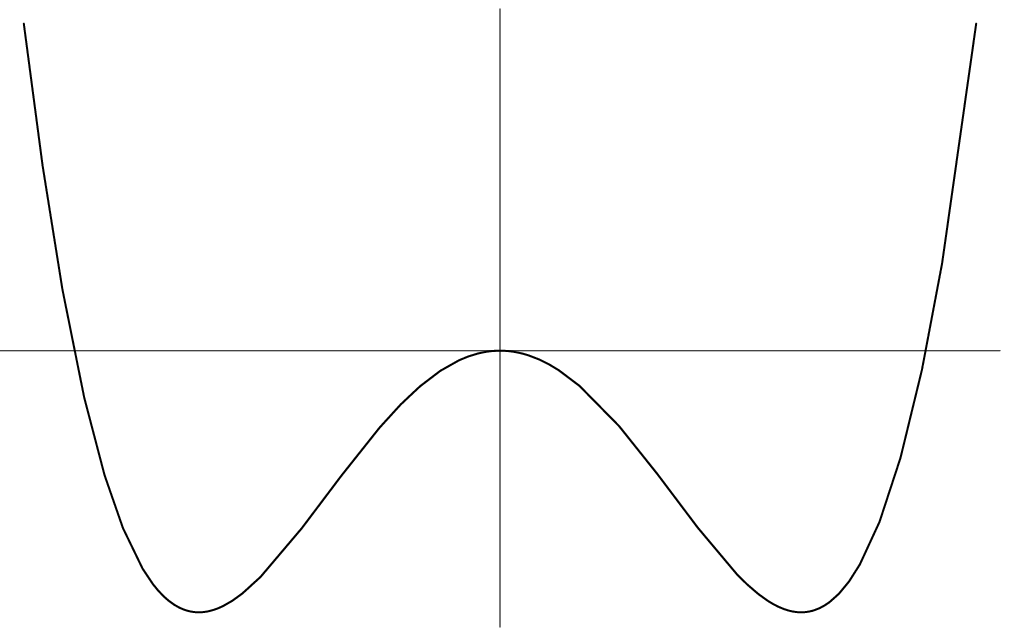,height=1.5in}} 
\caption{Potential of a single real scalar field responsible for
SSB} \label{2D Higgs potential plot}
\end{minipage}
\hspace{1cm}
\begin{minipage}{60mm}
\centerline{\psfig{figure=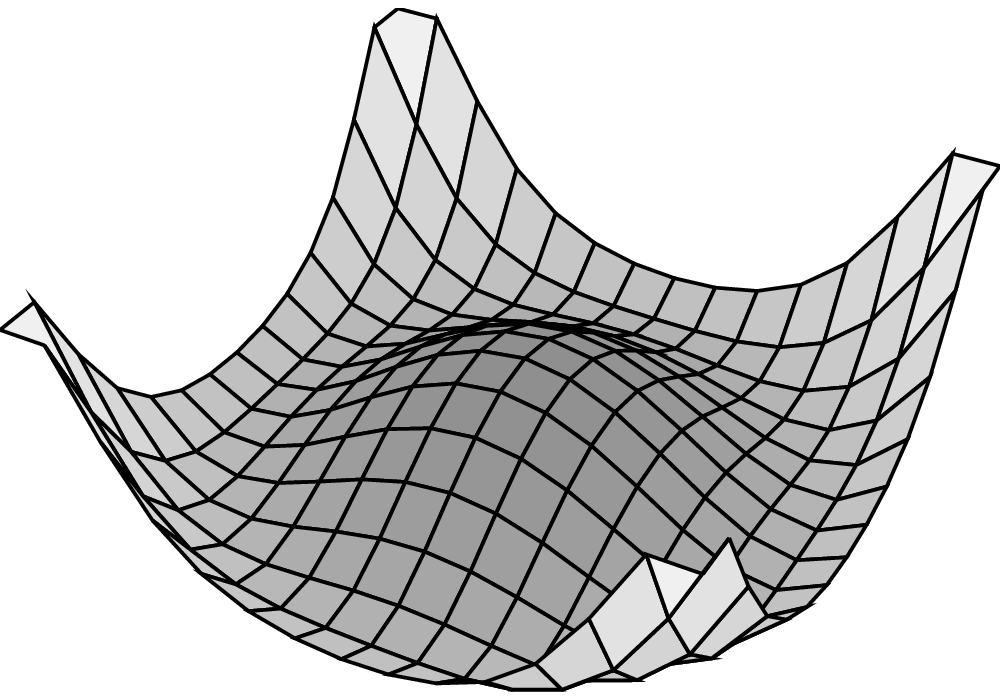,height=1.5in}}  
\caption{Potential of a complex scalar field responsible for SSB}
\label{3D Higgs potential plot}
\end{minipage}
\end{figure}

In physics, the standard way to achieve \emph{spontaneous symmetry
breaking} (SSB) is by introducing a scalar field\footnote{%
A famous example is the Higgs mechanism \cite{Higgs etc} in the
Standard Model to break the SU(2) $\times $U(1) gauge symmetry
spontaneously.} with a potential like
\begin{equation}
V(\Phi )=-\frac{1}{2}\mu ^{2}\Phi ^{2}+\frac{1}{4}\lambda \Phi
^{4}. \label{Higgs potential}
\end{equation}
Both the negative (mass)$^{2}$ $\Phi ^{2}$ term and $\lambda \Phi
^{4}$ term play a crucial role to make the potential behaving like
Figure \ref{2D Higgs potential plot} or Figure \ref{3D Higgs
potential plot} and possessing multiple true vacua. In the process
of choosing one among these true vacua, SSB is achieved and, in
the meanwhile, some topological defects like
domain walls or cosmic strings\footnote{%
If $\Phi $ is a single real scalar field, domain walls can form along with
SSB. On the other hand, if $\Phi $ denotes a complex scalar field or N $%
\left( \geq 2\right) $ real scalar fields, (cosmic) strings can form.} could
form.

The \emph{non-minimal coupling} between scalar fields and gravity
has been introduced in various topics in quantum field theories
and cosmology. In quantum field theories (for a comprehensive
review and a list of references, see the book
\cite{Buchbinder:1992rb}), the conformal invariance of the
massless scalar field with the non-minimal coupling constant $\xi
=1/6$ was first noted by Penrose \cite{Penrose:1964L} (see also
\cite{Chernikov:1968}). In the framework of quantum field theories
in curved space-time, the non-minimal coupling term can also be
introduced by the requirement of renormalizability
\cite{Buchbinder:1984xx} (for a review, see
\cite{Buchbinder:1989zz}). In cosmology (see \cite{Sahni:1998at}
\cite{Flachi:2000ca} and references therein), the usage of the
non-minimal coupling in the inflation models \cite{inflation
involving} and models for the cosmological constant problem
\cite{Sahni:1998at} \cite{Ford:1987de} (first proposed by Dolgov
\cite{Dolgov 1983}) has been widely explored. In addition, the
non-minimal coupling is also involved in an interesting model
called ``induced gravity'' which was first proposed by Zee in 1979
\cite{induced gravity} and is still explored nowadays (e.g. a
series of work about induced gravity and inflation by Kao
\cite{Kao induced gravity 2000}).

The role of the \emph{non-minimal coupling} in SSB and phase
transitions in 4D space-time has been widely discussed (for a
review, see \cite{Buchbinder:1992rb}). It has been pointed out
that the non-minimal coupling with the `external' gravitational
field may lead to SSB \cite{Janson1976&Grib1977}. It has also been
noted that SSB and phase transitions can be induced by curvature
via the non-minimal coupling with the `external' gravitational
field \cite{R-induced SSB}. Accompanying the quest of symmetry
breaking and vacuum stability in curved space-time
\cite{SB&VacuumStabi}, that the curvature can be a
symmetry-breaking factor has been shown. Note that, in the papers
mentioned above, the effect of the curved space-time is taken into
account by introducing an `external' gravitational field, which
means that the metric tensor of the space-time is treated as a
background, and the \emph{Ricci scalar} $\mathcal{R}$ in the
non-minimal coupling term is regarded as an `external' parameter
(like the temperature of a macroscopic system in thermodynamics).
This is different from our usage of the gravitational field to be
discussed below.

In this paper, we propose an alternative way to achieve SSB in 4D
and arbitrary higher ($D\geq 5$) dimensional space-time. We will
show that, by introducing the \emph{non-minimal coupling} between
the scalar field and gravity and keeping the negative (mass)$^{2}$
$\Phi ^{2}$ term, we can build up a new scenario to implement SSB
without using $\lambda \Phi ^{4}$ term. Avoiding the usage of
$\lambda \Phi ^{4}$ term could be a benefit when we consider SSB
in the higher dimensional space-time, because $\lambda \Phi ^{4} $
term is non-renormalizable in the formalism of quantum field
theories in $D\geq 5$ dimensional space-time. The gravitational
field introduced in this paper is not an `external' field or a
background field, but strongly depends on the scalar field through
the Einstein equations. Such a strong dependence plays a crucial
role in implementing SSB in our scenario. This is one of the
essential differences between the previous work mentioned in last
paragraph and our work in this paper.

To illustrate our idea specifically, we consider a simple case in
the third part of this paper: a world with an arbitrary space-time
dimension which is dominated by the `cosmological constant', the
`vacuum energy', and the `potential energy'\footnote{These three
terms---cosmological constant, vacuum energy, and potential
energy---can be summed up to form a single term which is treated
as an effective cosmological constant (or effective vacuum energy)
\cite{Zeldovich:1968} \cite{Weinberg:1989cp}: The energy-momentum
tensor of a vacuum in the quantum framework has the same form with
that of a cosmological constant, and hence makes a contribution to
the effective cosmological constant. In addition, the potential
part in the action of the scalar field $\Phi $ in the classical
framework also contributes a energy-momentum tensor which has the
same cosmological-constant form when $\Phi $ is constant in
space-time.\label{origin of effective lambda}}. (In fact, the
space-time dimension $D$ should be larger than two in our
discussion.) In this simple case, we will see that the potential
of the scalar field entails multiple true vacua, which is the key
element to implement SSB.

\section{Non-minimal coupling to gravity}

To introduce the non-minimal coupling between the scalar field and
gravity, we consider the Lagrangian density of the scalar field
$\Phi $ in $D$ dimensional space-time,
\begin{equation}
\mathcal{L}=\frac{1}{2}\mathcal{G}^{MN}\left( \partial _{M}\Phi \right)
\left( \partial _{N}\Phi \right) -V_{\mathcal{R}}\left( \Phi \right) ,%
\hspace{0.6cm}M,N=0,1,...,\left( D-1\right)
\end{equation}
where
\begin{equation}
V_{\mathcal{R}}\left( \Phi \right) =\frac{1}{2}\xi \mathcal{R}\Phi ^{2}-%
\frac{1}{2}\mu ^{2}\Phi ^{2}
\end{equation}
is the potential term for the scalar field $\Phi $ which includes
the non-minimal coupling to gravity via the \emph{Ricci scalar}
$\mathcal{R}$ with a coupling constant $\xi $ which is
\emph{positive}\footnote{Note that the convention for the metric
tensor in this paper is:
\begin{equation*}
\eta _{MN}=\left( 1,-1,-1,...,-1\right),
\end{equation*}
which is different from that in some of the papers mentioned in
`{\bf Introduction}' so that the sign of the non-minimal coupling
constant $\xi $ in this paper is opposite to that in them.} in our
consideration, and $ \mathcal{G}^{MN}$ is the metric tensor of the
$D$ dimensional space-time.

To discuss the behavior of the potential $V_{\mathcal{R}}$($\Phi
$), we need to know the behavior of the \emph{Ricci scalar}
$\mathcal{R}$, which can be obtained from the $D$ dimensional
\emph{Einstein equations}:
\begin{eqnarray}
G_{MN}&=& \Lambda \mathcal{G}_{MN} + \kappa T_{MN}, \quad \quad M,N=0,1,...,(D-1) %
          \nonumber \\
      &=& \kappa \left( \frac{1}{\kappa } \Lambda \mathcal{G}_{MN}+T_{MN}\right) %
          \nonumber \\
      &\equiv & \kappa \tilde{T}_{MN}, \label{effective energy momentum}
\end{eqnarray}
where $G_{MN}$ is the $D$ dimensional Einstein tensor, $\Lambda $
is the cosmological constant, $T_{MN}$ is the energy-momentum
tensor, $\tilde{T}_{MN}$ is the effective energy-momentum
tensor\footnote{The cosmological constant term $\Lambda
\mathcal{G}_{MN}$ can be absorbed into $T_{MN}$ (as shown in Eq.\
(\ref{effective energy momentum})) to form an effective
energy-momentum tensor $\tilde{T}_{MN}$ in which the contribution
from $\Lambda \mathcal{G}_{MN}$ is included.\label{effective
lambda}}, and $\kappa $ is a parameter (to be called ``$D$
dimensional gravitational constant'') which is positive and has
the same role with the Newton's constant $G$ in four dimensional
space-time. For simplicity, we consider the case for the perfect
fluid with the
effective energy-momentum tensor%
\begin{equation}
\tilde{T}_{\;\;\; N}^{M}=diag\left( \tilde{\rho}
,-\tilde{p}_{1},-\tilde{p}_{2},...,-\tilde{p}_{\left( D-1\right)
}\right) .
\end{equation}
By taking trace of the Einstein equations, it is straightforward
to get a relation between the \emph{Ricci scalar }$\mathcal{R}$
and $\rho ,p_{1},p_{2},...,p_{\left( D-1\right) }$:
\begin{equation}
-\frac{D-2}{2}\mathcal{R}=\kappa Tr\left( \tilde{T}_{\;\;\;
N}^{M}\right) =\kappa \left( \tilde{\rho}
-\tilde{p}_{1}-\tilde{p}_{2}-...-\tilde{p}_{\left( D-1\right)
}\right) .
\end{equation}

Consider a simple case that the effective energy-momentum tensor
is dominated by the effective cosmological constant $\Lambda
_{eff}$ which includes the original cosmological constant $\Lambda
$ and the possible contribution from the vacuum energy or the
potential energy (see footnote \ref{origin of effective lambda}):
\begin{equation}
\tilde{T}_{MN}\simeq \frac{1}{\kappa} \Lambda _{eff}
\mathcal{G}_{MN},
\end{equation}
and hence
\begin{equation}
\tilde{\rho} = \frac{1}{\kappa} \Lambda _{eff}
=-\tilde{p}_{1}=-\tilde{p}_{2}=...=-\tilde{p}_{\left( D-1\right)
}.
\end{equation}
Consequently, in this simple case,
\begin{equation}
\mathcal{R}=-\frac{2D}{D-2} \Lambda _{eff},
\end{equation}
which is positive(negative) for a negative(positive) effective
cosmological constant $\Lambda _{eff}$ and proportional to
$\Lambda _{eff}$.

The above result is essential in the following discussion about implementing
SSB in the third part of this paper. We will see that, if we treat the \emph{%
Ricci scalar} $\mathcal{R}$ as a parameter and write the potential as
\begin{equation}
V_{\mathcal{R}}\left( \Phi \right) =\frac{1}{2}\left( \xi \mathcal{R}-\mu
^{2}\right) \Phi ^{2},
\end{equation}
the ``coefficient'' $\left( \xi \mathcal{R}-\mu ^{2}\right) $ can
be negative for small $\Phi ^{2}$ and positive for large $\Phi
^{2}$ in some situation. And hence the potential
$V_{\mathcal{R}}\left( \Phi \right) $ behaves similarly to the
well-known potential (Eq.\ (\ref{Higgs potential})) for Higgs
mechanism, and entails multiple true vacua.

To see it more specifically, let us consider a simple case.

\section{A simple case to illustrate the way of implementing spontaneous symmetry breaking}

We consider the action in which only gravity and the scalar field
are introduced,
\begin{equation}
S=-\frac{1}{2\kappa }\int d^{D}x\sqrt{\mathcal{G}}\left( \mathcal{R}%
+2\Lambda \right) +\int d^{D}x\sqrt{\mathcal{G}}\mathcal{L},
\label{action}
\end{equation}
where
\begin{equation}
\mathcal{L}=\frac{1}{2}\mathcal{G}^{MN}\left( \partial _{M}\Phi \right)
\left( \partial _{N}\Phi \right) -V_{\mathcal{R}}\left( \Phi \right) ,%
\hspace{0.5cm}M,N=0,1,...,\left( D-1\right),
\end{equation}
\begin{equation}
V_{\mathcal{R}}\left( \Phi \right) =\frac{1}{2}\xi \mathcal{R}\Phi ^{2}-%
\frac{1}{2}\mu ^{2}\Phi ^{2}, \label{potential including
non-minimal coupling}
\end{equation}
$\kappa $ is the (positive) $D$ dimensional gravitational
constant, $\mathcal{G}$ is the absolute value of the determinant
of the metric tensor $\mathcal{G}_{MN}$, $\Lambda $ is the
cosmological constant, and $\xi $ is a \emph{positive} coupling
constant.

The variation of the above action with respect to the scalar field
$\Phi $ and the metric tensor $\mathcal{G}^{MN}$ yields the field
equation of $\Phi $ :
\begin{equation}
\Phi^{;J}_{\;\;\, ;J}+\left( \xi \mathcal{R}-\mu ^{2} \right) \Phi =0, %
\label{field equation of Phi}
\end{equation}
and the Einstein equations:
\begin{eqnarray}
G_{MN}&=&\Lambda \mathcal{G}_{MN} + \kappa \left\{ %
\left( \partial _{M}\Phi \right) \left( \partial _{N}\Phi \right)
- \mathcal{L}^{\left( 0 \right) }_{\Phi }\mathcal{G}_{MN} \right.
\nonumber \\ %
&& \quad \quad \quad \quad \quad \,\, %
\left. -\xi \Phi ^{2} G_{MN} + \xi \left( \Phi ^{2} \right) _{;M;N} %
- \xi \left( \Phi ^{2} \right)^{;J}_{\;\;\, ;J} \mathcal{G}_{MN}
\right\} , \label{Einstein equation}
\end{eqnarray}
where
\begin{equation}
\mathcal{L}^{\left( 0 \right) }_{\Phi } = %
\frac{1}{2} \mathcal{G}^{M'N'} %
\left( \partial _{M'} \Phi \right) %
\left( \partial _{N'} \Phi \right) + %
\frac{1}{2} \mu ^{2} \Phi ^{2}.
\end{equation}
(Note that there is no non-minimal coupling term in
$\mathcal{L}^{\left( 0 \right) }_{\Phi }$.) The semicolon `;'
denotes the `covariant derivative'. In the brace in Eq.\
(\ref{Einstein equation}), the term $-\xi \Phi ^{2}G_{MN}$, as
well as the terms $\xi \left( \Phi ^{2} \right) _{;M;N}$ and %
$-\xi \left( \Phi ^{2} \right)^{;J}_{\;\;\, ;J} \mathcal{G}_{MN}$,
is derived from the variation of the term $-\frac{1}{2}\xi
\mathcal{R}\Phi ^{2}\sqrt{\mathcal{G}}$ in the action. This term
modifies the Einstein equations, the gravitational constant
$\kappa $ and the cosmological constant $\Lambda $ as follows:
\begin{eqnarray}
G_{MN}&=&\Lambda ' \mathcal{G}_{MN} + \kappa _{eff} \left\{ %
\left( \partial _{M}\Phi \right) \left( \partial _{N}\Phi \right)
- \mathcal{L}^{\left( 0 \right) }_{\Phi }\mathcal{G}_{MN}
\right. \nonumber \\ %
&& \quad \quad \quad \quad \quad \quad \;\;\, %
\left. + \xi \left( \Phi ^{2} \right) _{;M;N} %
- \xi \left( \Phi ^{2} \right)^{;J}_{\;\;\, ;J} \mathcal{G}_{MN}
\right\}, \label{modified Einstein equation}
\end{eqnarray}
where the `modified cosmological constant' $\Lambda '$ and the
`effective gravitational constant' $\kappa _{eff}$ are defined as
\begin{eqnarray}
\Lambda '&\equiv &\frac{\Lambda }{1+\kappa \xi \Phi ^{2}}, \\
\kappa _{eff}&\equiv &\frac{\kappa }{1+\kappa \xi \Phi ^{2}}.
\end{eqnarray}
Note that both $\Lambda '$ and $\kappa _{eff}$ are not constant
parameters, but have the dependence on the square of the scalar
field. Taking trace of both sides of the \emph{modified} Einstein
equations (\ref{modified Einstein equation}) gives us the
\emph{Ricci scalar} $\mathcal{R}$ as a function of the scalar
field $\Phi $ and its covariant derivatives:
\begin{eqnarray}
\mathcal{R}&=& \frac{\kappa }{1 + \kappa \xi \Phi ^{2}} \left\{
-\frac{2D}{D-2} \left( \frac{1}{\kappa }\Lambda - \frac{1}{2} \mu
^{2} \Phi ^{2} \right) \right. \nonumber \\
&& \quad \quad \quad \quad \;\;\; \left. + \Phi ^{;J}\Phi _{;J} + %
2 \left( \frac{D-1}{D-2} \right) \xi %
\left( \Phi ^{2} \right) ^{;J}_{\;\;\, ;J} \right\}. %
\label{Ricci scalar}
\end{eqnarray}

For exploring the existence of multiple true vacua entailed by the
action (Eq.\ (\ref{action})) or the potential
$V_{\mathcal{R}}\left( \Phi \right) $ (Eq.\ (\ref{potential
including non-minimal coupling})), we need to find out the
constant solutions for the scalar field $\Phi $ and explore their
stability\footnote{A vacuum state is usually corresponding to a
stable constant-field solution.}. For constant $\Phi $, the field
equation of $\Phi $ (\ref{field equation of Phi}) and the
\emph{modified} Einstein equations (\ref{modified Einstein
equation}) become
\begin{equation}
\left( \xi \mathcal{R}-\mu ^{2} \right) \Phi = 0, %
\label{field equation for constant Phi}
\end{equation}
\begin{eqnarray}
G_{MN}&=& \Lambda ' \mathcal{G}_{MN}+ \kappa _{eff} \left\{
-\frac{1}{2}\mu ^{2}\Phi ^{2}\mathcal{G}_{MN} \right\} \nonumber \\
&=&\frac{\kappa }{1+\kappa \xi \Phi ^{2}}\left\{ \frac{1}{\kappa }%
\Lambda \mathcal{G}_{MN}-\frac{1}{2}\mu ^{2}\Phi
^{2}\mathcal{G}_{MN}\right\} \nonumber \\
&=&\kappa _{eff}V_{eff}\mathcal{G}_{MN}=\Lambda
_{eff}\mathcal{G}_{MN}, \label{Lambda dominated Einstein equation}
\end{eqnarray}
where the `\emph{effective} potential energy (or vacuum energy)'
$V_{eff}$ and the `\emph{effective} cosmological constant'
$\Lambda _{eff}$ are defined as
\begin{equation}
V_{eff}=\frac{1}{\kappa }\Lambda -\frac{1}{2}\mu ^{2}\Phi ^{2},
\end{equation}
\begin{equation}
\Lambda _{eff}=\frac{1}{1+\kappa \xi \Phi ^{2}} \left( \Lambda
-\frac{1}{2}\kappa \mu ^{2}\Phi ^{2}\right)=\kappa _{eff} V_{eff}.
\end{equation}
From Eq.\ (\ref{Lambda dominated Einstein equation}), we can see
that the constant-$\Phi $ solution is corresponding to a $\Lambda
$-dominated world, and hence the situation here becomes as simple
as the case mentioned in last section.

Taking trace of both sides of Eq.\ (\ref{Lambda dominated Einstein
equation}) (or applying the condition $\Phi =$ constant to Eq.\
(\ref{Ricci scalar})) gives us the \emph{Ricci scalar}
$\mathcal{R}$ as a function of the scalar field $\Phi $ :
\begin{equation}
\mathcal{R} = - \frac{2D}{\left( D-2 \right) } %
\frac{\kappa }{\left( 1+\kappa \xi \Phi ^{2} \right) } %
\left( \frac{1}{\kappa }\Lambda - \frac{1}{2}\mu ^{2}\Phi ^{2}\right) . %
\label{Ricci scalar for constant Phi}
\end{equation}
Using the above equation, Eq.\ (\ref{field equation for constant
Phi}) becomes
\begin{equation}
\left[ -\xi \frac{2D}{\left( D-2 \right) } \frac{\kappa }{\left(
1+\kappa \xi \Phi ^{2} \right) } \left( \frac{1}{\kappa }\Lambda -
\frac{1}{2}\mu ^{2}\Phi ^{2}\right) - \mu ^{2} \right] \Phi = 0 .
\end{equation}
Then we can obtain the constant solutions for the scalar field
$\Phi $ :
\begin{equation}
\Phi = \Phi _{\left( c \right) } = 0, \, \pm v
\end{equation}
where $v$ is defined by
\begin{equation}
v^{2} = \frac{D}{\kappa \xi } \left( \frac{D-2}{2D} + %
\frac{\xi \Lambda }{\mu ^{2}} \right)
\quad > 0 \quad \mbox{for} \quad %
\xi \Lambda > - \left( \frac{D-2}{2D} \right) \mu ^{2}. %
\label{vev}
\end{equation}
The corresponding metric tensor $\mathcal{G}_{MN}^{\left( c
\right) }$ (the solution(s) of the modified Einstein equations
corresponding to $\Phi = \Phi _{\left( c \right) }$ (Eq.\
(\ref{Lambda dominated Einstein equation}))) will be the metric
tensor of de Sitter or anti-de Sitter space-time, depending on the
`\emph{effective} cosmological constant' $\Lambda _{eff}$ is
positive or negative.

Now we need to explore the stability of these three constant-$\Phi
$ solutions ($\Phi = \Phi _{\left( c \right) }$,
$\mathcal{G}_{MN}=\mathcal{G}_{MN}^{\left( c \right) }$).
Considering the small variations around these solutions:
\begin{equation}
\left\{
\begin{array}{l}
\Phi = \Phi _{(c)} + \delta \Phi \\
\mathcal{G}_{MN} = \mathcal{G}_{MN}^{(c)} + \delta
\mathcal{G}_{MN} \;\; \longrightarrow \;\; %
\mathcal{R} = \mathcal{R}_{(c)} + \delta \mathcal{R}
\end{array}
\right. ,
\end{equation}
where $\mathcal{R}_{(c)}$ is the \emph{Ricci scalar} for $\Phi =
\Phi _{(c)}$, the field equation of $\Phi $ (\ref{field equation
of Phi}) becomes (up to $\mathcal{O}(\delta \Phi, \delta
\mathcal{R})$):
\begin{equation}
\left( \delta \Phi \right) ^{;J}_{\;\;\, ;J} %
+ \left( \xi \mathcal{R}_{(c)} - \mu ^{2} \right) \delta \Phi %
+ \xi \Phi _{(c)} \delta \mathcal{R} = 0, %
\label{field equation of delta Phi with R}
\end{equation}
while the equation for the \emph{Ricci scalar} $\mathcal{R}$
(\ref{Ricci scalar}) becomes (up to $\mathcal{O}(\delta \Phi,
\delta \mathcal{R})$):
\begin{eqnarray}
\delta \mathcal{R}&=& 0 \quad \mbox{for} \quad \Phi _{(c)}=0, %
\quad \mbox{or} \label{delta R at zero Phi} \\
\delta \mathcal{R}&=& 2\left( \frac{D-1}{D-2} \right) %
\left( \frac{\kappa \mu ^{2} \Phi _{(c)}}{1+\kappa \xi \Phi_{(c)}^{2}} %
\right) \delta \Phi %
+ 4 \left( \frac{D-1}{D-2} \right) \xi \Phi _{(c)} %
\left( \delta \Phi \right) ^{;J}_{\;\;\, ;J} \nonumber \\
&& \quad \quad \quad \quad \quad \quad \quad \quad \quad \quad
\quad \quad \quad \quad \quad \quad \quad \quad %
\mbox{for} \quad \Phi _{(c)}=\pm v. \label{delta R at vev}
\end{eqnarray}
Using above equations and Eq.\ (\ref{Ricci scalar for constant
Phi}), we can obtain the field equations of $\delta \Phi $ from
Eq.\ (\ref{field equation of delta Phi with R}): For $\Phi
_{(c)}=0$,
\begin{equation}
\left( \delta \Phi \right)^{;J}_{\;\;\, ;J} + \left[ - \left(
\frac{2D}{D-2} \xi \Lambda + \mu ^{2} \right) \right] \delta \Phi
= 0 , %
\label{field equation of delta Phi at zero Phi without R}
\end{equation}
where
\settowidth{\greater}{ >}
\begin{equation}
-\left( \frac{2D}{D-2} \xi \Lambda + \mu ^{2} \right) %
\mbox{ \raisebox{0.75ex}{$<$}\hspace{-0.95\greater}\raisebox{-0.75ex}{$>$} } %
0 \quad \mbox{for} \quad \xi \Lambda %
\mbox{ \raisebox{0.75ex}{$>$}\hspace{-0.95\greater}\raisebox{-0.75ex}{$<$} } %
- \left( \frac{D-2}{2D} \right) \mu ^{2}. %
\label{conditions for zero-Phi field equation}
\end{equation}
So the solution $\Phi =0$ is unstable or stable for $\xi \Lambda $
is greater or smaller than $-\left( \frac{D-2}{2D} \right) \mu
^{2}$. (Note that $\xi \Lambda > -\left( \frac{D-2}{2D} \right)
\mu ^{2}$ is the same condition with the one for $v^{2}>0$). On
the other hand, for $\Phi _{(c)}=\pm v$,
\begin{equation}
\left( \delta \Phi \right) ^{;J}_{\;\;\, ;J} %
+ 2 \left( \frac{D-1}{D-2} \right) \left[ 1 + 4 \left(
\frac{D-1}{D-2} \right) \xi ^{2} v^{2} \right] ^{-1} %
\left( \frac{\kappa \xi \mu ^{2} v^{2}}{1+\kappa \xi v^{2}}
\right) \delta \Phi = 0 ,
\end{equation}
where
\begin{equation}
2 \left( \frac{D-1}{D-2} \right) \left[ 1 + 4 \left(
\frac{D-1}{D-2} \right) \xi ^{2} v^{2} \right] ^{-1} %
\left( \frac{\kappa \xi \mu ^{2} v^{2}}{1+\kappa \xi v^{2}}
\right) > 0 \quad \mbox{for} \quad D > 2 . %
\label{condition for vev-Phi field equation}
\end{equation}
So the solutions $\Phi = \pm v$ are stable for $D>2$. (Remember
that, from Eq.\ (\ref{vev}), the condition $\xi \Lambda > -\left(
\frac{D-2}{2D} \right) \mu ^{2}$ should be fulfilled in order to
ensure the existence of the solutions $\Phi = \pm v$.)

Consequently we can conclude that, considering the space-time
dimension $D>2$, for $\xi \Lambda < -\left( \frac{D-2}{2D} \right)
\mu ^{2}$, there is only one constant solution $\Phi = 0$, which
is stable and corresponding to one true vacuum; while for $\xi
\Lambda > -\left( \frac{D-2}{2D} \right) \mu ^{2}$, there are one
unstable constant solution $\Phi = 0$ and two stable constant
solutions $\Phi = \pm v$ which are corresponding to two true
vacua.

\section{Discussion}

We have shown that the potential $V_{\mathcal{R}}\left( \Phi
\right) $ does entail multiple true vacua and hence be able to
result in spontaneous symmetry breaking under suitable conditions.
In particular, we can see, from Eq.\ (\ref{vev})(\ref{conditions
for zero-Phi field equation})(\ref{condition for vev-Phi field
equation}), that the conditions
\begin{equation}
\left\{ %
\begin{array}{ccl}
             D &>& 2 \\
        \kappa &>& 0 \\
           \xi &>& 0 \\
   \xi \Lambda &>& -\left( \frac{D-2}{2D} \right) \mu ^{2}
\end{array}
\right.
\end{equation}
should be fulfilled in order to produce spontaneous symmetry
breaking in our scenario.

In our scenario, the gravitational constant $\kappa $ will undergo
a ``phase transition'' accompanying the spontaneous symmetry
breaking: %
\settowidth{\back}{>}
\begin{equation*}
\kappa \quad %
\mbox{\raisebox{0.9ex}{\underline{\ \ \ \ SSB\ \ \ \ }}
\hspace{-2\back}$>$} %
\quad \kappa _{eff}=\frac{\kappa }{1+\kappa \xi v^{2}}
\end{equation*}
when the scalar field $\Phi $ rolls down to one of the true vacua
from the origin. It is interesting to see the large $\xi \Lambda $
limit of $\kappa _{eff}$:
\begin{equation*}
\xi \Lambda \gg \mu ^{2}, \quad \quad \kappa _{eff} \simeq \left(
\frac{\mu ^2}{D \xi \Lambda} \right) \kappa \ll \kappa .
\end{equation*}
Accordingly, even though the original gravitational constant
$\kappa $ could be arbitrarily large, we still can get a small
effective gravitational constant $\kappa _{eff}$ by choosing (or
tuning) the parameters: $\mu $, $\xi $, and $\Lambda $ such that
the fraction $\mu ^{2}$/$\xi \Lambda $ is small enough. This
phenomenon is similar to the result of the ``induced gravity''
model \cite{induced gravity}.

It would be of interest to find, numerically, the solutions in our
model, that is, the solutions of the coupled equations
--- the field equation of the scalar field $\Phi $ (\ref{field equation of Phi})
and the modified Einstein equations (\ref{modified Einstein
equation})\footnote{In fact, there is one redundant equation among
them.}. Except the trivial solutions: $\Phi =\Phi _{(c)}$,
$\mathcal{G}_{MN}=\mathcal{G}_{MN}^{(c)}$, which have been
discussed in last section, we want to find out non-trivial
solutions, especially the domain-wall solution(s) for studying the
domain-wall formation under our new SSB scenario and the
relationship between such kind of domain wall and the
Randall-Sundrum scenario \cite{Randall:1999vf}
\cite{Randall:1999ee} in 5D space-time. In our preliminary sight
on it, we do see that the configuration of this domain wall is
similar to the configuration of the brane world in the
Randall-Sundrum scenario \cite{Randall:1999vf}, in which the
appropriate ``Friedmann equation(s)'' (describing the expansion or
contraction of the brane world) can be obtained successfully
\cite{expanding R-S scenario}. We also wish to explore more
details on it.

\section*{Acknowledgements}

This work is supported in part by the National Science Council,
Taiwan, R.O.C. (NSC 89-2112-M-002 062) and by the CosPA project of
the Ministry of Education (MOE 89-N-FA01-1-4-3).



\newpage

\begin{thebibliography}{99}

\bibitem{Higgs etc}
P.~W.~Higgs,
Phys.\ Lett.\ {\bf 12} (1964) 132;
%
P.~W.~Higgs,
Phys.\ Rev.\ {\bf 145} (1966) 1156;
%
T.~W.~Kibble,
Phys.\ Rev.\ {\bf 155} (1967) 1554.

\bibitem{Buchbinder:1992rb}
I.~L.~Buchbinder, S.~D.~Odintsov and I.~L.~Shapiro,
\emph{Effective Action in Quantum Gravity} %
(IOP Pub., Bristol, UK, 1992).

\bibitem{Penrose:1964L}
R.~Penrose, %
in \emph{Relativity, Groups and Topology}, %
edited by C.~DeWitt and B.~S.~DeWitt %
(Gordon and Breach, New York, 1964).

\bibitem{Chernikov:1968}
N.~A.~Chernikov and E.~A.~Tagirov, %
Ann.\ Inst.\ H.\ Poincare {\bf 9A} (1968) 109.

\bibitem{Buchbinder:1984xx}
I.~L.~Buchbinder,
Sov.\ Phys.\ J.\ {\bf 27} (1984) 1054.

\bibitem{Buchbinder:1989zz}
I.~L.~Buchbinder, S.~D.~Odintsov and I.~L.~Shapiro,
Riv.\ Nuovo Cim.\ {\bf 12N10} (1989) 1.


\bibitem{Sahni:1998at}
V.~Sahni and S.~Habib,
Phys.\ Rev.\ Lett.\ {\bf 81} (1998) 1766 [hep-ph/9808204].

\bibitem{Flachi:2000ca}
A.~Flachi and D.~J.~Toms,
Phys.\ Lett.\ B {\bf 478} (2000) 280 [hep-th/0003008].


\bibitem{inflation involving}
R.~Fakir and W.~G.~Unruh,
Phys.\ Rev.\ D {\bf 41} (1990) 1783;
%
D.~S.~Salopek, J.~R.~Bond and J.~M.~Bardeen,
Phys.\ Rev.\ D {\bf 40} (1989) 1753;
%
E.~W.~Kolb, D.~S.~Salopek and M.~S.~Turner,
Phys.\ Rev.\ D {\bf 42} (1990) 3925;
%
R.~Fakir, S.~Habib and W.~Unruh,
Astrophys.\ J. {\bf 394} (1992) 396;
%
R.~Fakir and S.~Habib,
Mod.\ Phys.\ Lett.\ A {\bf 8} (1993) 2827;
%
M.~S.~Turner and L.~M.~Widrow,
Phys.\ Rev.\ D {\bf 37} (1988) 3428;
%
R.~Fakir and W.~G.~Unruh,
Phys.\ Rev.\ D {\bf 41} (1990) 1792.


\bibitem{Ford:1987de}
L.~H.~Ford,
Phys.\ Rev.\ D {\bf 35} (1987) 2339.

\bibitem{Dolgov 1983}
A.~D.~Dolgov,
in \emph{The Very Early Universe}, %
edited by G.~W.~Gibbons, S.~W.~Hawking and S.~T.~C.~Siklos
(Cambridge University Press, Cambridge, 1983)

\bibitem{induced gravity}
A.~Zee,
Phys.\ Rev.\ Lett.\ {\bf 42} (1979) 417;
%
L.~Smolin,
Nucl.\ Phys.\ B {\bf 160} (1979) 253;
%
S.~L.~Adler,
Rev.\ Mod.\ Phys.\ {\bf 54} (1982) 729.


\bibitem{Kao induced gravity 2000}
W.~F.~Kao,
Phys.\ Rev.\ D {\bf 62} (2000) 084009 [hep-th/0003153];
%
{\bf 62} (2000) 087301 [hep-th/0003206];
%
{\bf 61} (2000) 047501.


\bibitem{Janson1976&Grib1977}
M.~M.~Janson,
Lett.\ Nuovo Cim.\ {\bf 15} (1976) 231;
%
A.~A.~Grib and V.~M.~Mostepanenko,
Pisma Zh.\ Eksp.\ Teor.\ Fiz.\ {\bf 25} (1977) 302 [JETP Lett.\
{\bf 25} (1977) 277].


\bibitem{R-induced SSB}
G.~M.~Shore, Ann.\ Phys.\ {\bf 128} (1980) 376;
%
B.~Allen,
Nucl.\ Phys.\ B {\bf 226} (1983) 228;
%
K.~Ishikawa,
Phys.\ Rev.\ D {\bf 28} (1983) 2445;
%
I.~L.~Buchbinder and S.~D.~Odintsov,
Class.\ Quant.\ Grav.\ {\bf 2} (1985) 721;
%
Yad.\ Fiz.\ {\bf 42} (1985) 1268;
%
J.~Perez-Mercader,
Phys.\ Lett.\ B {\bf 223} (1989) 300.


\bibitem{SB&VacuumStabi}
D.~J.~Toms,
Phys.\ Rev.\ D {\bf 25} (1982) 2536;
%
L.~H.~Ford and D.~J.~Toms,
Phys.\ Rev.\ D {\bf 25} (1982) 1510;
%
A.~Vilenkin and L.~H.~Ford,
Phys.\ Rev.\ D {\bf 26} (1982) 1231;
%
A.~Vilenkin,
Nucl.\ Phys.\ B {\bf 226} (1983) 504;
%
V.~N.~Melnikov and S.~V.~Orlov,
Phys.\ Lett.\ A {\bf 70} (1979) 263.



\bibitem{Zeldovich:1968}
Y.~B.~Zeldovich,
Sov.\ Phys.\ - Uspekhi {\bf 11} (1968) 381.

\bibitem{Weinberg:1989cp}
S.~Weinberg,
Rev.\ Mod.\ Phys.\ {\bf 61} (1989) 1.


\bibitem{Randall:1999vf}
L.~Randall and R.~Sundrum,
Phys.\ Rev.\ Lett.\ {\bf 83} (1999) 4690 [hep-th/9906064].

\bibitem{Randall:1999ee}
L.~Randall and R.~Sundrum,
Phys.\ Rev.\ Lett.\ {\bf 83} (1999) 3370 [hep-ph/9905221].

\bibitem{expanding R-S scenario}
E.~E.~Flanagan, S.~H.~Tye and I.~Wasserman,
Phys.\ Rev.\ D {\bf 62} (2000) 044039 [hep-ph/9910498];
%
P.~Kraus,
JHEP{\bf 9912} (1999) 011 [hep-th/9910149];
%
C.~Csaki, M.~Graesser, C.~Kolda and J.~Terning,
Phys.\ Lett.\ B {\bf 462} (1999) 34 [hep-ph/9906513].


\end{thebibliography}
\end{document}